\documentclass[fleqn,usenatbib]{mnras}

\usepackage[utf8]{inputenc}

\usepackage{newtxtext}

\usepackage[T1]{fontenc}
\usepackage{lmodern}

\DeclareRobustCommand{\VAN}[3]{#2}
\let\VANthebibliography\thebibliography
\def\thebibliography{\DeclareRobustCommand{\VAN}[3]{##3}\VANthebibliography}

\usepackage{graphicx}	
\usepackage{amsmath}	
\usepackage{amssymb}	
\usepackage{xcolor}
\usepackage{booktabs}   
\usepackage{rotating}   
\usepackage{caption}    
\usepackage[version=4]{mhchem} 

\title[Disequilibrium assumption in retrieval]{A parameterised approach to disequilibrium retrievals in the JWST era: Application to NIRCam observations of HD 189733b}

\author[Taylor et al]{
Jake Taylor$^{1}$\thanks{E-mail: jake.taylor@physics.ox.ac.uk},
Shang-Min Tsai$^{2,3}$,
Vivien Parmentier$^{4}$,
Chloe Fisher$^{1}$
$\&$ Michael Line$^{5}$
\\
$^{1}$Department of Physics, University of Oxford, Parks Rd, Oxford, OX1 3PU, UK\\
$^{2}$Department of Earth and Planetary Sciences, University of California, Riverside, CA, USA \\
$^{3}${Institute of Astronomy \& Astrophysics, Academia Sinica (ASIAA), Taipei, Taiwan}\\
$^{4}$Universit\'e C\^ote d'Azur, Observatoire de la C\^ote d'Azur, CNRS, Laboratoire Lagrange, French Riviera, France \\
$^{5}$School of Earth and Space Exploration, Arizona State University, 781 South Terrace Road, Tempe, AZ 85281, USA}

\date{Accepted XXX. Received YYY; in original form ZZZ}

\pubyear{2025}

\begin{document}
\label{firstpage}
\pagerange{\pageref{firstpage}--\pageref{lastpage}}
\maketitle

\begin{abstract}
Atmospheric retrievals are a widely used technique for inferring the physical and chemical properties of exoplanetary atmospheres from observed spectra. A common simplifying assumption in such analyses is that the atmosphere is in thermochemical equilibrium, which allows the use of precomputed chemical abundance grids as a function of pressure, temperature, metallicity ([M/H]), and carbon-to-oxygen ratio (C/O). However, exoplanet atmospheres often deviate from equilibrium, particularly at lower temperatures or in the presence of strong vertical mixing. In this work, we investigate the impact of disequilibrium chemistry on retrieval outcomes by generating synthetic James Webb Space Telescope (JWST) observations of HD\,189733\,b with varying strengths of vertical mixing. We demonstrate that assuming thermochemical equilibrium can lead to significant biases in the retrieved atmospheric parameters, including incorrect estimates of C/O and [M/H]. To address this, we incorporate transport-induced quenching of carbon and nitrogen-bearing species into the retrieval framework by allowing the quench pressures to be free parameters. We show that this approach recovers the correct bulk atmospheric properties in most cases. Finally, we apply our disequilibrium retrieval model to published JWST/NIRCam transmission observations of HD\,189733\,b and find tentative evidence for quenching. We also find tentative evidence for the photochemically active region of the atmosphere via a newly developed H$_2$S parameterisation, this is the first time this has been constrained in a hot Jupiter atmosphere.
\end{abstract}




\section{Introduction}

Exoplanet spectra can teach us a lot about the chemical composition and the thermal structure of exoplanet atmospheres \citep[e.g.][]{Stevenson2014,Parmentier2026}. However, extracting accurate information from spectra is a complex, often degenerate process \citep[e.g.][]{Welbanks2019}. The inverse problem is typically solved using Bayesian inference methods, which explore the parameter space by iteratively evaluating a forward model to constrain the range of atmospheric properties consistent with the observations. As such, all parameter inferences depend fundamentally on the forward model used and, particularly, how the thermo-chemical structure of the atmosphere is parameterised \citep[e.g.][]{Line2013}. As of now, three main families of models are used in the current literature for atmospheric retrieval.

The first family of models are called the "free-chemistry" models, whereby both chemical abundances and thermal profiles are free to vary independently. Because of the large number of parameters, this approach often assumes that chemical abundances are constant with height in the atmosphere. Their advantage is that the relative chemical abundances are not tied to any pre-conception we might have about chemistry in planetary atmospheres and can therefore extract information we may not have anticipated. One example of this is from the detection of SO$_2$ in the atmosphere of WASP-39b \citep{Powell2024,Tsai2023}. However, given their large number of free parameters, these free chemistry methods are prone to overfitting, leading to possible biased atmospheric detections.

A second family of models similarly considers the thermal profile as a set of free parameters, but assumes that the atmosphere is in chemical equilibrium, meaning that only a few parameters, such as the metallicity ([M/H]) and the carbon-to-oxygen ratio (C/O) can be used to parameterise the chemistry instead of a large number of individual molecules. Typically, this is through coupling a chemical-kinetics code with a radiative transfer framework \citep{Al-Refaie2022,Taylor2023} In such a framework, the molecular abundances become directly tied to the thermal structure and can vary significantly with altitude. However, whereas chemical equilibrium is likely a good approximation for hot Jupiters and similarly irradiated exoplanets \citep{Venot2015}, it would likely lead to biased conclusions for cooler objects where disequilibrium chemistry becomes important \citep{Miles2020}.

A last family of models are models that calculate the thermal profile of the atmosphere through radiative/convective equilibrium models coupled to kinetic models for the chemistry \citep[e.g.][]{Mang2026}. These models are the most consistent, as both chemistry and temperature are solved together. A key point of these models is that they are using our knowledge of physics and chemistry to link the observable atmosphere to the deep, usually unobservable, thermochemical structure of the planet. Indeed, vertical mixing, often parameterised with the diffusion coefficient, can quench chemical species to their values in the deep atmospheric layers, where chemical equilibrium holds true. As a consequence, atmospheric retrievals through a grid of radiative-convective-disequilibrium chemistry models can provide insights into the deep thermal structure of the atmosphere that free retrievals would be incapable of doing \citep{Welbanks2024,Sing2024}. However, despite the much lower number of possible free parameters, these models require more time to converge and must rely on a pre-computed grid of models that are then interpolated during the retrieval process \citep{Kreidberg2015,Fu2024}. Furthermore, they assume that the thermal profile is determined by the radiative/convective equilibrium, an assumption that can be erroneous when atmospheric dynamics can transport energy from the dayside to the nightside of the planet \citep{Wiser2026}.

\begin{figure*}
    \centering
    \includegraphics[width=0.99\textwidth]{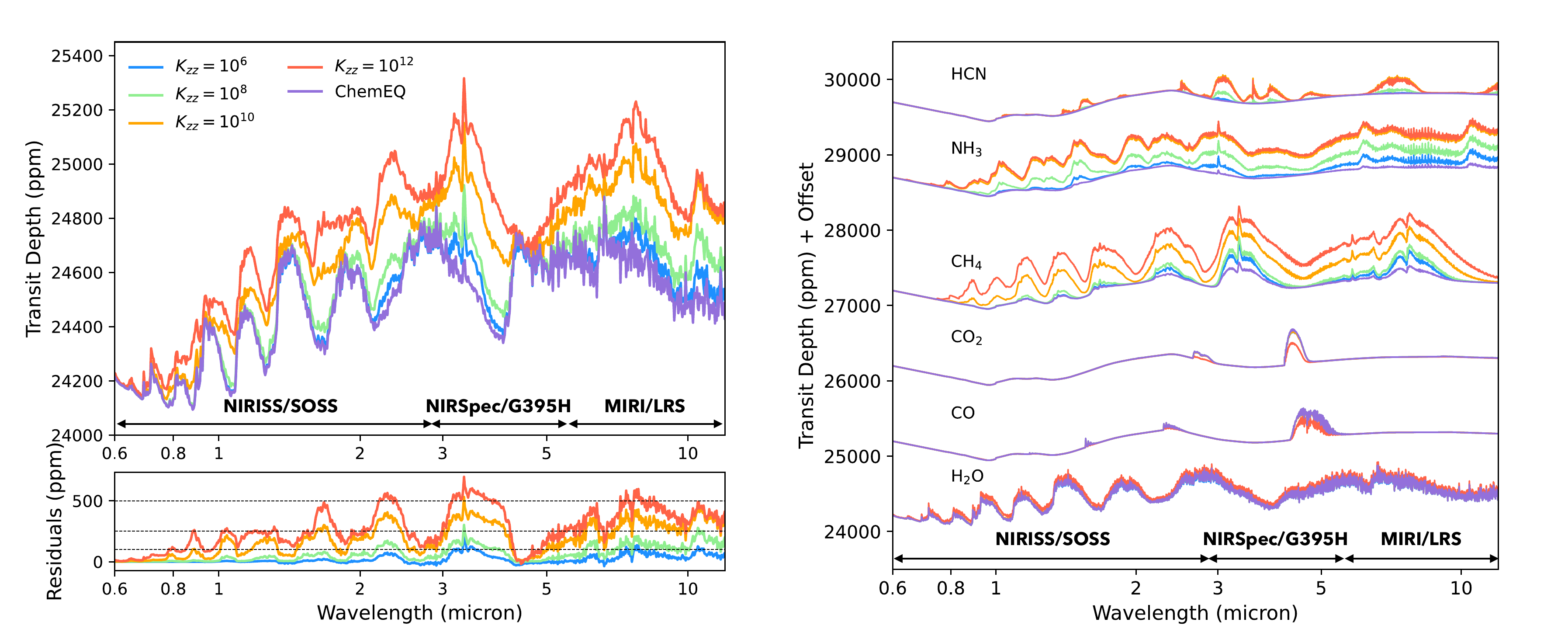}
    \caption{Left: Model transmission spectra of a hot Jupiter with system parameters corresponding to HD 189733b. The spectra were generated by post-processing chemical abundance profiles from \texttt{VULCAN} using \texttt{NemesisPy}. Five atmospheric models are shown: one assuming chemical equilibrium, and four incorporating varying strengths of vertical mixing, parameterised by different eddy diffusion coefficients ($K_{\text{zz}}$). The lower panel shows the spectral residuals between the equilibrium model and each non-equilibrium case. Horizontal dashed lines mark deviations of 100, 250, and 500 ppm for reference. Right: Contribution of individual molecular species to the transmission spectrum as a function of $K_{\text{zz}}$, illustrating how vertical mixing affects the spectral signatures of key absorbers. We have indicated the approximate wavelength coverage for NIRISS/SOSS, NIRSpec/G395H, and MIRI/LRS at the bottom of each plot.}
    \label{fig:vulcan_models}
\end{figure*}

In this study, we present an alternative framework that tries to solve the shortcomings of the three previously described approaches. We focus on modelling transport-induced quenching, an effect more readily incorporated into retrieval frameworks. Specifically, we examine how atmospheric mixing processes, such as eddy diffusion and global-scale advection, can drive chemical abundances out of equilibrium \citep{Cooper2006, Moses2011, Parmentier2013, Drummond2020, Tsai2017, Zamyatina2023, Tsai2024, ZhangZ2025}. These processes can lead to the vertical homogenisation of species. In our framework, we retain the flexibility of the thermal profile parameterisation while parameterising the possible effects of disequilibrium chemistry with a parameterised quenching approach. This allows us to explore a wide variety of possible planetary thermal structures, like in free or chemical equilibrium retrievals, while retaining the important link between thermal structure, transport, and chemistry.

The layout of this paper is as follows: in Section \ref{Methods}, we present our model and new parameterisation that can account for disequilibrium chemistry from vertical mixing. In Section \ref{Results}, we present the results from a range of synthetic observations, where we compare chemical equilibrium retrievals to our disequilibrium retrievals. In Section \ref{JWST}, we apply our disequilibrium framework to published JWST observations, and finally, we present our conclusions in Section \ref{Conclusions}.

\section{Methods}
\label{Methods}
In this section, we outline our model and the synthetic datasets generated to assess the robustness of our model.

\begin{figure*}
    \centering
    \includegraphics[width=\textwidth]{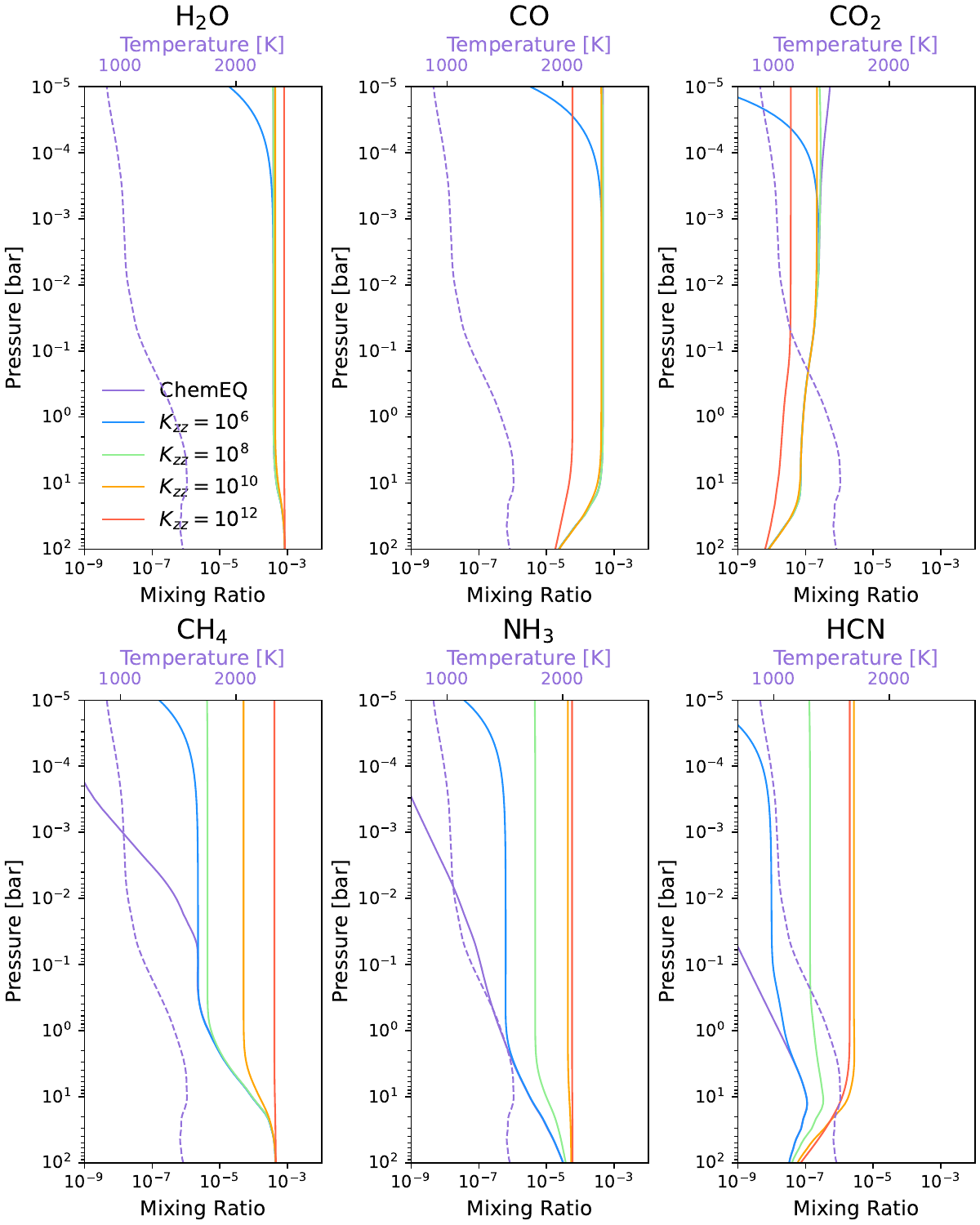}
    \caption{The volume mixing ratios of the key species investigated in this study. Each panel shows the volume mixing ratio for a given molecule and multiple $K_{zz}$ values. We plot the input TP profile with a dashed purple line. The metallicity and C/O ratios assumed in these models are solar.}
    \label{fig:VMRs}
\end{figure*}

\subsection{Model setup}
\label{setup}

\texttt{NEMESIS} is a radiative transfer and retrieval tool originally developed to study planetary atmospheres within the solar system \citep{Irwin2008}, but extensively adapted to study exoplanet atmospheres \citep[e.g.][]{Barstow2017,Taylor2023}. The framework is now fully Pythonised, and we use this version of the code for this analysis \citep[called \texttt{NemesisPy} from here on out][]{Yang2024}. \texttt{NemesisPy} uses the correlated-k method to compute the molecular and atomic opacities \citep{Lacis1991} and nested sampling, specifically \texttt{PyMultiNest} to sample the parameter-space \citep{buchner14,feroz09}. For our simulations, we consider the following opacity sources: H$_2$O \citep{Polyansky2018}, CO \citep{Li2015}, CO$_2$ \citep{Yurchenko2020}, CH$_4$ \citep{Yurchenko2017}, NH$_3$ \citep{Coles2019}, SO$_2$ \citep{Underwood2016}, H$_2$S \citep{Azzam2016}, HCN \citep{Barber2014} and C$_2$H$_2$ \citep{Chubb2020} \footnote{For the quenching retrievals, no sulfur chemistry is used}. We also model the collisional induced absorption from H$_2$-H$_2$ and H$_2$-He \citep{Borysow1989,Borysow1989b,Borysow1990,Borysow1997,Borysow2001,Borysow2002}. The gas opacities are computed using k-tables with resolution R=1000, obtained from the ExoMol database \citep{Chubb2021}, before being channel averaged to the resolution of the observations. We model chemical equilibrium calculations by using the \texttt{FastChem} \citep{Kitzmann2024} package, which has been coupled to \texttt{NemesisPy} to obtain metallicity and C/O values from JWST observations \citep{Banerjee2024}. 

We adopt the pressure–temperature (P–T) profile from \citet{Moses2011} and use \texttt{VULCAN} \citep{Tsai2017, Tsai2021} to generate chemical abundance profiles for HD\,189733\,b under varying strengths of vertical mixing, parameterised by $K_{\mathrm{zz}}$. This allows us to explore how transport-induced disequilibrium can alter atmospheric composition. Photochemistry is not included in this study. We employ the \texttt{VULCAN} NCHO chemical network\footnote{\url{https://github.com/exoclime/VULCAN/blob/master/thermo/NCHO_photo_network.txt}} to simulate the full N–C–H–O chemistry, but restrict our analysis to the following species: H$_2$O, CH$_4$, CO, CO$_2$, HCN, NH$_3$, and C$_2$H$_2$. We restrict to these molecules as they have the largest VMRs and spectral impact over the wavelengths considered.

The chemical abundance profiles from \texttt{VULCAN} are post-processed with \texttt{NemesisPy} to produce transmission spectra, which are shown in Fig.~\ref{fig:vulcan_models}. All models assume solar C/O and metallicity, following \citet{Lodders2009}. We consider four values of $K_{\mathrm{zz}}$ = $10^{6}$, $10^{8}$, $10^{10}$, and $10^{12}$cm$^2$s$^{-1}$, spanning a range of plausible mixing strengths for hot Jupiters \citep{Parmentier2013, Komacek2019, Moses2022}.

The left panel of Fig.~\ref{fig:vulcan_models} shows the full transmission spectra used to generate the synthetic observations for this study. Notably, even at the lowest mixing strength, the CH$_4$ feature at $\sim$3.3$\mu$m deviates by more than 100 ppm from the equilibrium model. The right panel presents the contribution of individual molecular species to the spectra, highlighting which molecules are most affected by vertical mixing and identifying the spectral regions where these differences are most apparent.

In Fig.~\ref{fig:VMRs}, we present the vertical volume mixing ratios (VMRs) for key molecular species that significantly affect the transmission spectrum. These profiles illustrate how the atmospheric composition varies with the assumed strength of vertical mixing. The dominant carbon- and oxygen-bearing species in the atmosphere of HD\,189733\,b, H$_2$O, CO, and CO$_2$, exhibit nearly identical vertical profiles between 10$^3$ and 10$^{-3}$ bar across all but the most extreme case of vertical mixing ($K{\mathrm{zz}} = 10^{12}$cm$^2$s$^{-1}$), where deviations become apparent. The profiles in Fig.~\ref{fig:VMRs} demonstrate that, for hot Jupiters, H$_2$O, CO, and CO$_2$ are largely unaffected by vertical mixing within the pressure range most relevant to transmission spectroscopy. Consequently, their spectral features show minimal variation between models, as also evident in Fig.~\ref{fig:vulcan_models}.

The equilibrium abundances of CH$_4$, NH$_3$, and HCN exhibit a strong pressure dependence as a function of $K{\mathrm{zz}}$, indicating a high sensitivity to vertical mixing. These species are particularly affected by quenching, with their abundances becoming fixed at pressures where the vertical transport timescale becomes shorter than the chemical timescale. We do not apply our quenching framework to HCN, as its abundance is primarily influenced by photochemical pathways rather than thermochemical equilibrium. As such, it is not amenable to the simplified quenching approximation employed in this study \citep{Tsai2021}. For all species, the weakest mixing case ($K_{\mathrm{zz}} = 10^{6}$cm$^2$s$^{-1}$) results in a characteristic exponential decline in VMRs at pressures lower than 10$^{-3}$ bar, where chemical equilibrium begins to dominate due to slower dynamical transport.

We consider three different JWST observing configurations: a single transit with NIRISS/SOSS and NIRSpec/G395H individually, and a combined NIRISS/SOSS and NIRSpec/G395H observation. These are typical observing configurations used to maximise the inference about the atmospheres of exoplanets. We omit MIRI/LRS due to the low precision. Synthetic observations for each mode are generated using \texttt{PandExo~2.0} \citep{Batalha2017, Batalha2022}. {We assume a noise floor of 10 ppm and a resolution of R=100. We note that HD 189733 is too bright for these observing modes, and these simulations serve as a "JWST-like" case. In fact, it has been shown that PandExo estimates the uncertainty on the observations to be more precise than seen in observations \citep{Alderson2025}. The average uncertainty on the NIRISS/SOSS data is around 18 ppm, and the average uncertainty on the NIRSpec/G395H observations is around 25ppm. For each observing configuration, we construct five synthetic datasets: one based on the chemical equilibrium assumption, and four based on the varying vertical mixing strengths ($K_{\mathrm{zz}}$). Each dataset is then analysed using three retrieval approaches: one assuming thermochemical equilibrium, and the others allowing for disequilibrium chemistry by fitting for quench pressures (elemental and molecular). In total, this results in 15 retrievals per observing configuration.

To compute the atmospheric temperature structure, we adopt the parameterisation described by \citet{Madhusudhan2009} (referred to as MS09). This approach divides the atmosphere into three layers: an upper region where temperature inversions are not permitted, a middle region where an inversion is allowed, and a deep layer that is assumed to be isothermal. The parameterisation introduces six free parameters in total. The prior distributions for all free parameters are listed in Table~\ref{tab:priors}. Furthermore, we perform another set of simulations that assumes a temperature structure described by \citet{Parmentier2014} (referred to as PG14), which is a semi-analytical model governed by five free parameters.

To explore the parameter space, we couple our parametric forward model with the Bayesian nested sampling algorithm \textsc{PyMultiNest} \citep{feroz09, buchner14}. We adopt 500 live points for all simulations and set the evidence tolerance to 0.5, which is the default value in \textsc{PyMultiNest} and ensures robust convergence of the retrievals.

\subsection{Disequilibrium parametrisation}
\label{sec:diseq_param}
To incorporate disequilibrium chemistry into our framework, we manipulate the vertical chemical abundance profiles output from \texttt{FastChem~3.0} by introducing free parameters which fit the quenching pressure in the atmosphere. We do this in two ways: "elemental", where we quench molecules with similar elemental constituents, and "molecular", where we quench the individual molecules. These parameters represent the pressure levels at which the abundances of relevant species are assumed to become vertically homogenised due to transport-induced quenching.

This "elemental" approach is motivated by the dominant chemical pathways in carbon and nitrogen chemistry. The primary carbon-bearing species are CH$_4$ and CO, linked through the net reaction:
\begin{equation}
\text{CH}_4 + \text{H}_2\text{O} \rightarrow \text{CO} + 3\text{H}_2,
\end{equation}
and therefore, H$_2$O is also quenched at the same pressure as the carbon species.

Similarly, the dominant nitrogen-bearing species are NH$_3$ and N$_2$, connected via the net reaction:
\begin{equation}
2\text{NH}_3 \leftrightarrow \text{N}_2 + 3\text{H}_2.
\end{equation}

We do not include opacity from N$_2$ in our simulations, as it lacks significant spectral features in the wavelength ranges considered. Therefore, the molecular species governed by the quenching parameterisation are H$_2$O, CO, CO$_2$, CH$_4$, and NH$_3$.

We treat carbon and nitrogen-bearing species separately, allowing for independent quench pressures ($\log P_{\rm q,C}$ and $\log P_{\rm q,N}$), as these groups typically quench at different pressure levels \citep{Moses2014, Tsai2018}. For a given species, the vertical volume mixing ratio is modified such that the abundance is fixed to its equilibrium value at the specified quench pressure. This can be expressed as:

\begin{equation}
\label{eq:quench}
X(P) =
\begin{cases}
X_{\mathrm{eq}}(P), & P > P_{\mathrm{quench}} \\[6pt]
X_{\mathrm{eq}}\bigl(P_{\mathrm{quench}}\bigr), & P \le P_{\mathrm{quench}}
\end{cases}
\end{equation}

where $X(P)$ is the volume mixing ratio at pressure $P$, $X_{\mathrm{eq}}(P)$ is the equilibrium volume mixing ratio computed from \texttt{FastChem~3.0}, and $P_{\mathrm{quench}}$ is the retrieved quench pressure for the relevant species. That is, above the quench pressure (i.e., at lower pressures), the abundance is held constant at the value computed at $P_{\mathrm{quench}}$, while deeper in the atmosphere (at higher pressures), the abundance follows thermochemical equilibrium. This method can be applied to any set of chemical profiles, whether generated via precomputed grids or on-the-fly computations.

We extend this framework by quenching individual molecules, rather than grouping them by their elements. We therefore quench CH$_4$ and NH$_3$ following the same technique as shown in Eq. \ref{eq:quench}. We want to explore if a targeted molecular quenching framework can perform similarly to, or outperform, the elemental quenching framework, where we quench carbon species and nitrogen species together.

We note that quench pressure parameterisations have been employed in previous studies \citep[e.g.][]{Morley2017, Molliere2020, Benneke2024}; however, these typically assumed a single quench pressure for all species. Other approaches have attempted to retrieve the eddy diffusion coefficient $K_{\mathrm{zz}}$ directly \citep[e.g.][]{Kawashima2021, Al-Refaie2022}. While such methods are grounded in chemical kinetics, they also propagate uncertainties associated with reaction rate data and timescale assumptions. By contrast, our quench pressure framework is model-independent and agnostic to chemical kinetics, providing a flexible and computationally efficient means of accounting for disequilibrium chemistry. We also note that \citet{Wogan2025} showed the vertical profile of CO$_2$ is quenched twice; however, this would be relevant for colder objects, whereas this study focuses purely on the impact of transport-induced quenching in hot Jupiter atmospheres \citep{Tsai2018}.

\begin{table}
    \centering
    \begin{tabular}{cccc}
    \hline
         Parameter & Unit & Prior Distribution & Limits \\
         \hline
         \multicolumn{4}{c}{\citet{Madhusudhan2009} TP profile} \\
         \hline
         $T_0$ & K & Uniform & $[400,2000]$\\
         $\log(P_1)$ & bar & Log-uniform & $[-6,2]$\\
         $\log(P_2)$ & bar & Log-uniform & $[-6,2]$\\
         $\log(P_3)$ & bar & Log-uniform & $[-2,2]$\\
         $a_1$ & -- & Uniform & $[0.02,2]$ \\
         $a_2$ & -- & Uniform & $[0.02,2]$ \\
         \hline
         \multicolumn{4}{c}{\citet{Parmentier2014} TP profile} \\
         \hline
         $\log(\kappa_{\rm IR})$ & -- & Uniform & $[-4,0]$\\
         $\log(\gamma_1)$ & -- & Uniform & $[-4,4]$\\
         $\log(\gamma_2)$ & -- & Uniform & $[-4,4]$\\
         $T_{\rm irr}$ & K & Uniform & $[500,2000]$\\
         $\alpha$ & -- & Uniform & $[0,1]$\\
         \hline
         \multicolumn{4}{c}{Atmospheric composition and radius} \\
         \hline
         $\log({\rm Met})$ & -- & Log-uniform & $[-1,2]$ \\
         C/O & -- & Uniform & $[0,1]$ \\
         $xR_{\rm p}$ & -- & Uniform & $[0.8,1.2]$ \\
         \hline
         \multicolumn{4}{c}{Quench pressure parameters} \\
         \hline
         $\log P_{\rm q,C}$ & bar & Log-uniform & $[-5.5,2]$ \\
         $\log P_{\rm q,N}$ & bar & Log-uniform & $[-5.5,2]$ \\
         \hline 
    \end{tabular}
    \caption{Prior distributions for all possible model parameters for \texttt{NemesisPy}.}
    \label{tab:priors}
\end{table}

\section{Results}
\label{Results}

In this section, we present the results of our retrieval simulations. We examine the performance of individual JWST instrument modes in retrieving bulk atmospheric properties under three different assumptions: (i) chemical equilibrium, (ii) our elemental quenching framework, and (iii) our molecular quenching framework. We then explore how the retrieval outcomes are affected when multiple instrument modes are combined.

\subsection{Individual Instruments}
\begin{figure*}
    \centering
    \includegraphics[width=0.99\linewidth]{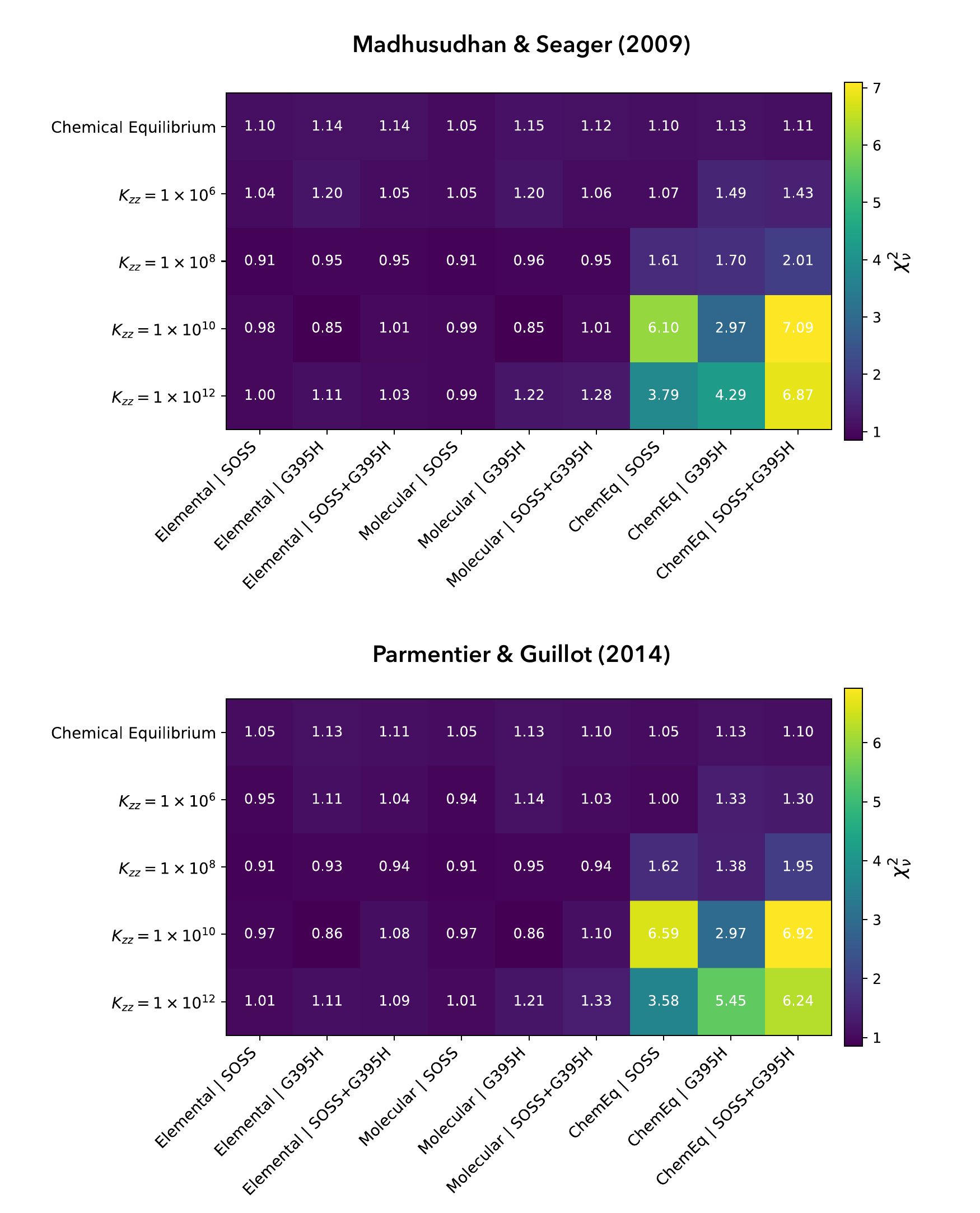}
    \caption{Heatmap of the reduced $\chi^2$ values for all of the simulations described in Section \ref{setup}. We label each panel with the value of the reduced $\chi^2$ and colour-code with the corresponding colour-bar for easier visual inspection. We have split into two heatmaps, one for each TP parameterisation assumed.}
    \label{fig:heatmap}
\end{figure*}
We begin by examining the atmospheric constraints achievable using individual JWST instrument modes: NIRISS/SOSS and NIRSpec/G395H. Each instrument has distinct strengths and limitations, driven by differences in spectral coverage and resolving power.

NIRISS/SOSS covers wavelengths from 0.6 to 2.8 $\mu$m with a resolving power of $R \sim 700$. This range is well-suited to probing the optical and near-infrared spectrum, enabling the detection of alkali metals such as Na and K, as well as molecules including H$_2$O and CH$_4$ \citep[e.g.][]{Fu2022, Radica2023, Holmberg2023}. NIRSpec/G395H spans 2.8 to 5.2 $\mu$m with a higher resolving power of $R \sim 2700$. It complements the shorter wavelength by covering more carbon-bearing features, as well as sulfur-bearing species \citep{Alderson2023, August2023, Moran2023}.

We present a goodness-of-fit heatmap in Figure \ref{fig:heatmap} to demonstrate how well each model assumption fits the data. It can be seen that as soon as we introduce quenching into our observations, the fit by the chemical equilibrium framework begins to get worse with increasing $K_{\mathrm{zz}}$. In contrast, the elemental and molecular quenching frameworks provide reasonable fits to the observations for all mixing strengths. We can conclude that traditional chemical equilibrium modelling cannot adequately explain observations of hot Jupiters that have detectable vertical mixing.

In Figures~\ref{fig:CtoO_Violin_Plot} and \ref{fig:MtoH_Violin_Plot}, we present the retrieved values of C/O and [M/H] for each instrument mode, respectively. The horizontal dashed lines indicate the true (input) values used to generate the synthetic observations. Retrieval results are shown as violin plots, which illustrate the full posterior distributions. The median value is marked with a point, and the associated 1$\sigma$ uncertainty is shown as an error bar. The top row of panels is the result when assuming an MS09 TP profile, and the bottom row of panels is the result when assuming a PG14 TP profile.

We display in purple the disequilibrium retrievals using the elemental quenching framework (quenching carbon species and nitrogen species), red for the disequilibrium retrievals using the molecular quenching framework (quenching individual molecules), and blue for the chemical equilibrium retrievals. Each panel corresponds to a different atmospheric mixing scenario, ranging from equilibrium to a highly mixed atmosphere with $K_{\mathrm{zz}} = 10^{12}$cm$^2$s$^{-1}$.

We first focus on the left-most panels in Figures~\ref{fig:CtoO_Violin_Plot} and \ref{fig:MtoH_Violin_Plot}, which are the results of the chemical equilibrium retrievals. A similar trend is seen regardless of the assumed TP profile. We can see that, for NIRISS/SOSS alone, we constrain a C/O ratio and [M/H] that is smaller than the input. We attribute this to the retrieval only being able to use the H$_2$O bands to determine the overall C/O and [M/H], as CH$_4$ is negligible in the equilibrium setup. The H$_2$O bands in themselves can be degenerate for a wide range of C/O and [M/H] values in the absence of other absorbers \citep{Feinstein2023}. In contrast, we can recover the input C/O and [M/H] for the NIRSpec/G395H simulation. It can be seen that we can recover the same results for the chemical equilibrium and disequilibrium models, demonstrating that if there is no detectable vertical mixing, the model will correctly recover the chemical equilibrium case. 

We now analyse the retrieval results for atmospheres that include vertical mixing. Once again, a similar trend is seen regardless of the TP profile assumed, with few exceptions. The chemical equilibrium retrievals (blue violin plots) struggle to recover the true input values of C/O and [M/H] as the mixing strength increases. For NIRISS/SOSS, there are no cases in the MS09 assumption where the input C/O is correctly recovered to 1-$\sigma$. For the PG14 assumption, only the C/O value of the $K_{\mathrm{zz}} = 10^{8}$ is recovered correctly. The metallicity is only correctly recovered for the lowest mixing strength in the MS09 models ($K_{\mathrm{zz}} = 10^6$\,cm$^2$\,s$^{-1}$), but deviates significantly for stronger mixing. In no case is the metallicity recovered for the PG14 models. For NIRSpec/G395H, the chemical equilibrium retrieval fails to recover the input C/O and [M/H] for the MS09 models. This is the same as the C/O ratios recovered for the PG14 models; however, the metallicity is correctly recovered for the mixing strengths $K_{\mathrm{zz}} = 10^6$ and $10^8$ \,cm$^2$\,s$^{-1}$. We therefore highlight the breakdown of the chemical equilibrium assumption when vertical mixing is present. 

The disequilibrium retrievals (purple and red violin plots) perform significantly better in recovering the input C/O and [M/H] values. Both MS09 and PG14 obtain similar constraints. For NIRISS/SOSS, the retrieved values are consistent with the true values for most cases. The molecular quenching has C/O values that deviate from the input for the mixing values of $K_{\mathrm{zz}} = 10^6$ and $10^8$\,cm$^2$\,s$^{-1}$, and [M/H] for $K_{\mathrm{zz}} = 10^{12}$. The elemental quenching deviates from the [M/H] in the $K_{\mathrm{zz}} = 10^{12}$ case. Similar to NIRISS/SOSS, the retrieved values for NIRSpec/G395H are consistent with the true values for most cases. With only one deviation from the input value, which is seen by the molecular quench simulation for $K_{\mathrm{zz}} = 10^{12}$. We can conclude that the elemental or molecular quenching frameworks provide a more accurate representation of the bulk properties in the atmosphere compared to the chemical equilibrium assumption.

\subsection{Combining Instrument Modes}
\begin{figure*}
    \centering
    \includegraphics[width=0.99\textwidth]{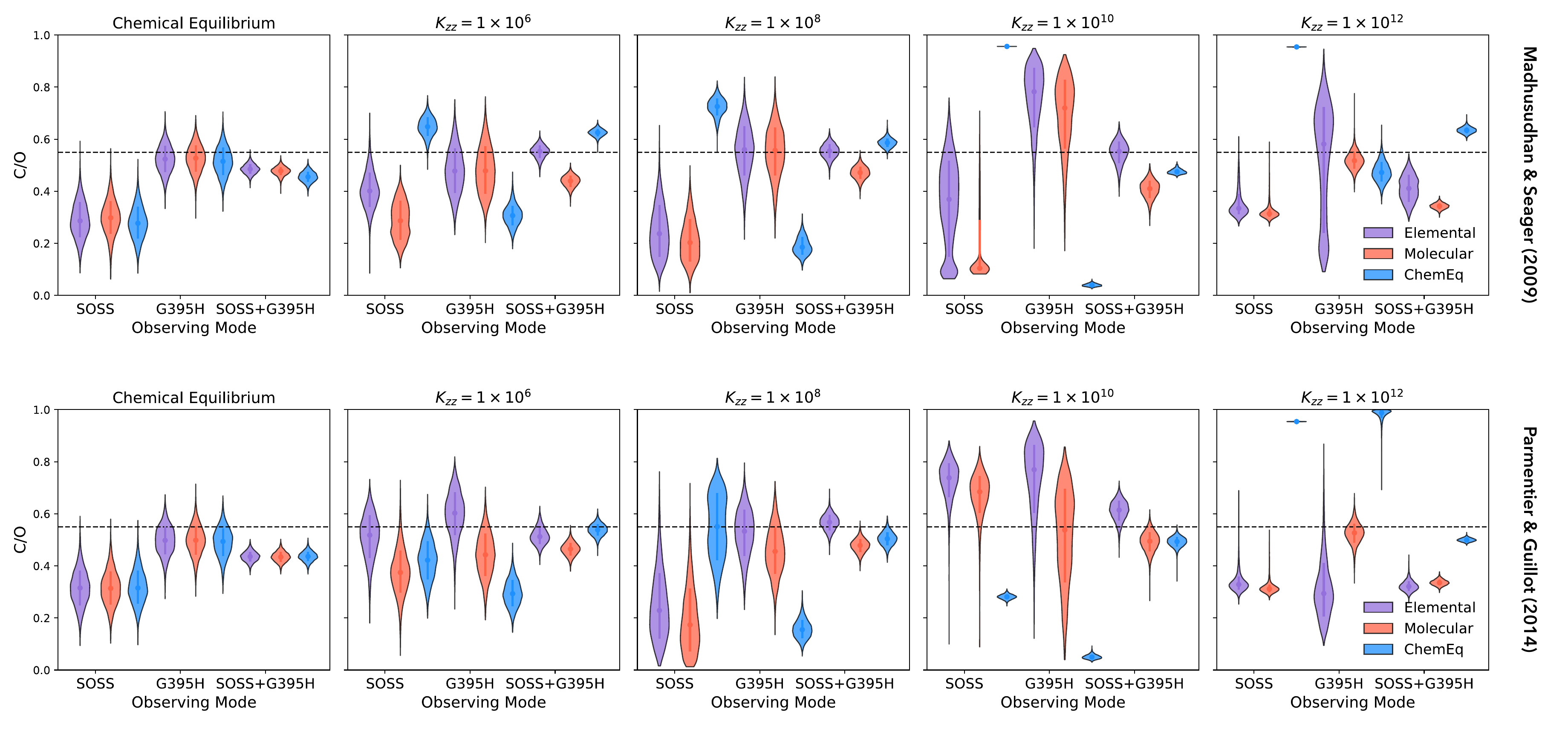}
    \caption{Violin plots of the retrieved C/O values for the different instrument modes. The top panel shows the retrieved C/O values when assuming a \citet{Madhusudhan2009} TP profile, the bottom panel shows the retrieved C/O values when assuming a \citet{Parmentier2014} TP profile. The horizontal dashed lines represent the true value of C/O. The error bars represent the 1-$\sigma$ uncertainties. The colours represent the model assumption: purple for elemental quenching, red for molecular quenching, and blue for chemical equilibrium.}
    \label{fig:CtoO_Violin_Plot}
\end{figure*}

\begin{figure*}
    \centering
    \includegraphics[width=0.99\textwidth]{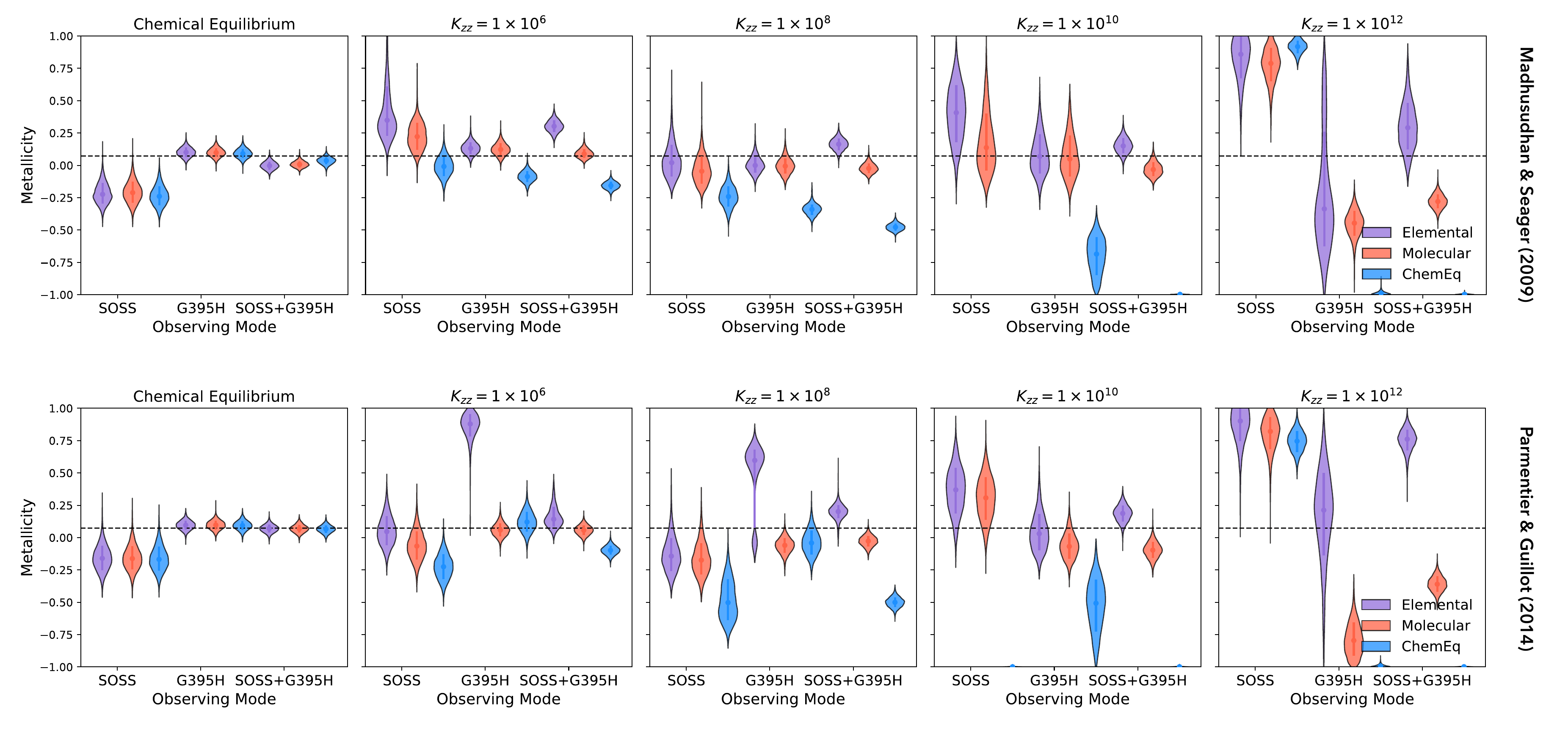}
    \caption{Violin plots of the retrieved [M/H] values for the different instrument modes. The top panel shows the retrieved [M/H] values when assuming a \citet{Madhusudhan2009} TP profile, the bottom panel shows the retrieved C/O values when assuming a \citet{Parmentier2014} TP profile. The horizontal dashed lines represent the true value of C/O. The error bars represent the 1-$\sigma$ uncertainties. The colours represent the model assumption: purple for elemental quenching, red for molecular quenching, and blue for chemical equilibrium.}
    \label{fig:MtoH_Violin_Plot}
\end{figure*}

A typical observation setup is to combine NIRISS/SOSS observations with NIRSpec/G395H observations, in order to probe a wide range of molecular features. We explore how our results change when doing this. The retrieved C/O and [M/H] values for these combined cases are presented as violin plots in Fig.~\ref{fig:CtoO_Violin_Plot} and ~\ref{fig:MtoH_Violin_Plot} respectively.

We find a significantly tighter posterior distribution compared to the individual cases, as well as consistency between MS09 and PG14, although agreement with the true values is not always improved. Firstly, the chemical equilibrium results are similar to those of the individual cases, with the recovered values having significant variations from the input values. For the elemental quenching retrievals, the violin plots encapsulate the input values in nearly all cases. The exceptions are the [M/H] value for $K_{\mathrm{zz}} = 10^{6}$ and the C/O value for $K_{\mathrm{zz}} = 10^{12}$. The molecular quenching struggles more, recovering systematically lower C/O ratios in all cases, potentially driven by the preference for low C/O values for the NIRISS/SOSS data. The [M/H] is recovered for all cases apart from $K_{\mathrm{zz}} = 10^{12}$.

\subsection{Retrieved Quenching Pressures}

Allowing the quenching pressures to vary as free parameters in the retrieval provides a means of inferring the vertical mixing strength present in the atmosphere. To assess the effectiveness of this approach, we compare the retrieved quenching pressures with the approximate values obtained from the \texttt{VULCAN} models, as shown in Fig.~\ref{fig:quenching_pressures}. The retrieved quenching pressures for elemental nitrogen and carbon are shown in purple and blue. The retrieved quenching pressures for molecular ammonia and methane are shown in red and yellow. As with the C/O and [M/H] retrievals, we use violin plots to represent the posterior distributions, with horizontal dashed lines indicating the true values. The approximate values for the carbon quenching are denoted with a triangle, and the approximate value for the nitrogen quenching is denoted with a star. To determine the quench pressure, we compute the variation of the volume mixing ratio between two layers, and if the difference is sufficiently small, we conclude that it is quenched. We determine this value with the following equation:
\begin{equation}
    a > \frac{X_{z+1} - X_z}{X_z}
\end{equation}
where $X_z$ is the volume mixing ratio in layer $z$, and $a$ is the value we consider the layer difference to be quenched. For carbon species, we assume $a = 0.25$, and for nitrogen species, we use $a = 0.1$. We require different values for $a$ because the vertical equilibrium profile of NH$_3$ varies less compared to CH$_4$.

We find that nitrogen chemistry quenches at deeper atmospheric levels than carbon chemistry, consistent with theoretical expectations as nitrogen conversion proceeds more slowly than carbon \citep{Moses2011}. This is why it is important to fit the quench pressure of nitrogen and carbon species individually. We find that the quench pressures are recovered in most cases. However, the carbon quenching pressure is recovered at higher pressures in the atmosphere compared to the true values. We note that for vertical mixing of $K_{\mathrm{zz}} = 10^{10}$ and $K_{\mathrm{zz}} = 10^{12}$, nitrogen is quenching around 100 bar, which is the bottom of the atmosphere in our retrieval model. These simulations demonstrate that, if vertical mixing is present, we are able to recover the approximate pressure levels at which quenching is occurring, which tells us about the chemistry deeper in the atmosphere. We find a consistent trend between MS09 and PG14, although the posterior distributions for PG14 are much wider. The broader posteriors likely reflect stronger degeneracies between the PG14 temperature-profile parameters, which propagate into the retrieved chemical parameters.

\begin{figure*}
    \centering
    \includegraphics[width=0.99\textwidth]{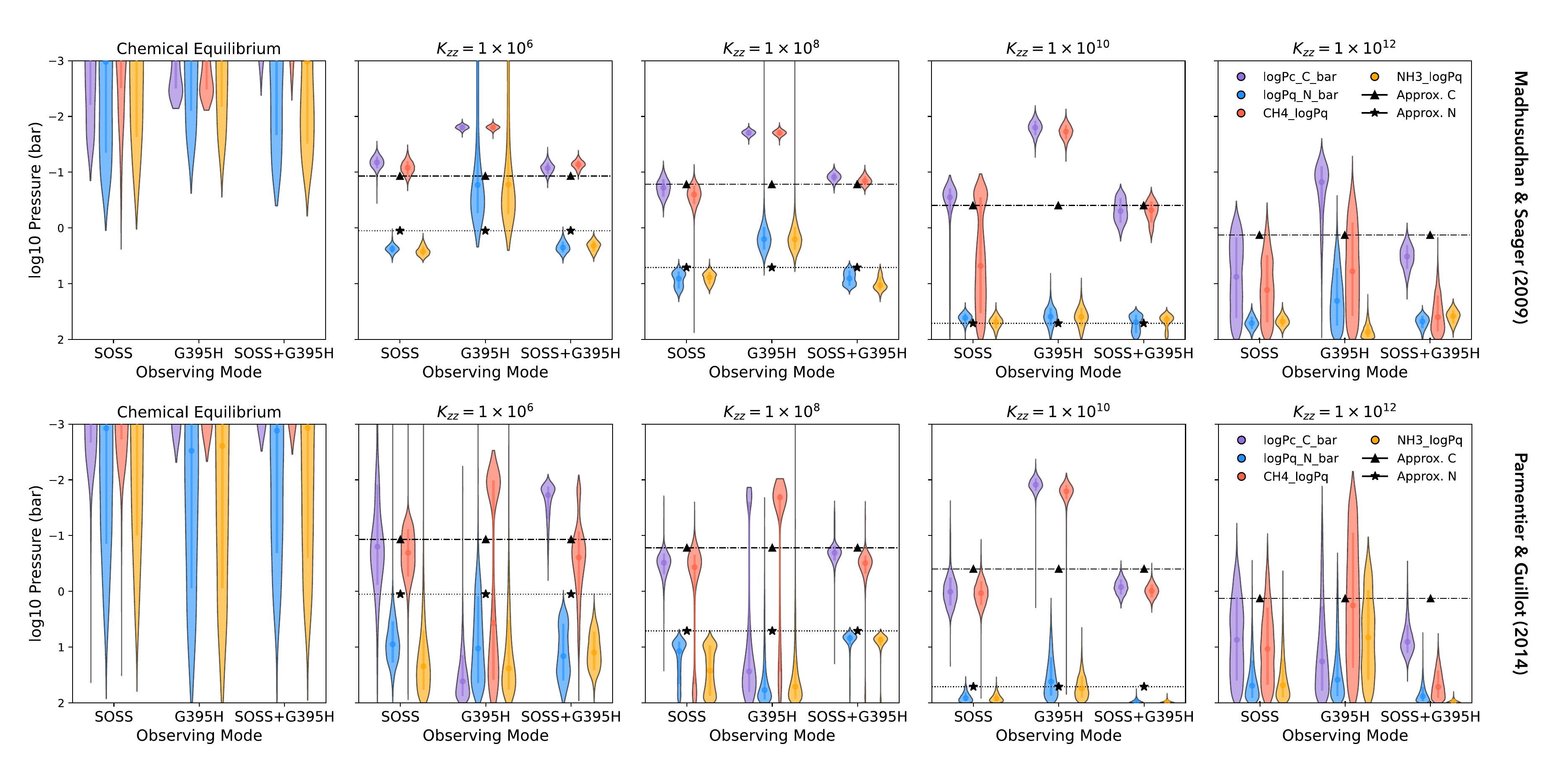}
    \caption{The retrieved quenching pressures from the elemental and molecular quenching simulations. The colour of the violin plots corresponds to the quench parameter being retrieved, with purple, blue, red, and yellow being the elemental carbon, elemental nitrogen, CH$_4$, and NH$_3$, respectively. The horizontal dashed lines represent the true values of the quenching pressures, with the triangle for the approximate carbon quenching pressure and the star for the approximate nitrogen quenching pressure. Within the violin plot, we represent the median value with error bars representing the 1-$\sigma$ uncertainties. The top set of panels are the simulations assuming the \citet{Madhusudhan2009} TP profile, and the bottom panel assume the \citet{Parmentier2014} TP profile.} 
    \label{fig:quenching_pressures}
\end{figure*}

\section{Application to JWST Observations of HD\,189733\,b}
\label{JWST}

In this section, we apply our quenching retrieval framework to recent JWST transmission observations of HD\,189733\,b. This planet was observed as part of the Guaranteed Time Observations (GTO) programme 1633, using the NIRCam grism mode with the F444W and F322W2 filters. These observations cover the spectral range from 2.4 to 5.2 $\mu$m, which is broadly similar to the range probed by NIRSpec/G395H. However, the NIRCam configuration extends further into the shorter-wavelength portion of the H$_2$O absorption band.

The JWST observations were analysed by \citet{Fu2024}, who performed four independent data reductions, each yielding consistent transmission spectra. Owing to this consistency, they conducted their atmospheric retrievals using one of the reduced datasets. Their retrievals were performed with \texttt{CHIMERA} under a free-chemistry approach in which the volume mixing ratio of each molecule is assumed to be constant with altitude and fitted independently.

They employed a non-isothermal temperature structure following the parameterisation of \citet{Parmentier2014}, and included inhomogeneous cloud coverage in the model. The atmospheric composition was assumed to be H$_2$-dominated, with the following species considered: H$_2$O, CO, CO$_2$, NH$_3$, HCN, SO$_2$, C$_2$H$_2$, and H$_2$S. From their retrieval, they constrained the abundances of H$_2$O, CO, CO$_2$, and H$_2$S, while the remaining species were unconstrained, most notably CH$_4$, which appeared to be depleted relative to expectations from chemical equilibrium.

A second set of retrievals was performed by \citet{Fu2024} using a grid of 1D radiative-convective photo-equilibrium (1D-RCPE) models, generated with the \texttt{CHIMERA} forward model coupled to \texttt{VULCAN} to incorporate photochemical processes. From their 1D-RCPE modelling, they concluded that the atmosphere of HD 189733b has a C/O ratio between 0.1 and 0.14, and a super-stellar metallicity of approximately 3–5$\times$ solar. These results are primarily driven by the retrieved abundances of H$_2$O and CO$_2$, and the non-detection of SO$_2$.

Their modelling favours a 1D-RCPE profile with an internal temperature ($T_{\rm int}$) of 500\,K as providing the best fit to the data (see Fig.~2 of \citealt{Fu2024}). Interestingly, they find that the abundance of H$_2$S required to best match the observed spectrum is roughly 10$\times$ solar, significantly higher than expected given the metallicity constraints inferred from the relative feature strengths of H$_2$O and CO$_2$. 

Although Bayes factors and statistical confidence levels were not reported for the free-chemistry retrievals, they were calculated for the 1D-RCPE analysis. In this framework, H$_2$O and CO$_2$ are detected at greater than 10$\sigma$ confidence, CO at 5$\sigma$, and H$_2$S at 4.5$\sigma$, representing the first reported detection of H$_2$S in an exoplanet atmosphere.

An independent analysis of the observations was conducted by \citet{Zhang2025}, who performed a separate reduction of the JWST transmission data, incorporated JWST emission observations, and extended the study to include all available pre-JWST measurements. Their analysis of the JWST transmission and eclipse spectra is broadly consistent with that of \citet{Fu2024}, concluding that methane depletion is required to explain the observed features. However, they did not explicitly model disequilibrium chemistry, noting in their conclusions that it may play an important role in interpreting JWST observations.

\subsection{Retrieval and H$_2$S Parameterisation}
\label{sec:h2s_param}
We pose the following question: can our quenching framework provide a physical explanation for the observed depletion of CH$_4$, and the observed sulfur constraints, seen in the atmosphere of HD\,189733\,b? To investigate this, we take the transmission spectrum published by \citet{Fu2024} and perform a suite of retrievals.

Across the models, all models include an inflation term for each dataset (as described in \citet{Burningham2017}), as well as a cloud top pressure, a Rayleigh enhancement factor, and a scattering slope, as defined in \citep{Taylor2023}. 

We note that our quenching framework ("DisEq``) does not take into account sulfur chemistry, specifically the interaction between H$_2$S and SO$_2$ due to photochemistry. To account for this, we develop a model that parameterises the VMR of H$_2$S with a deep abundance and a single break pressure, above which the abundance follows a monotonic power-law decline. Specifically, the vertical profile is:

\[
X_{\mathrm{H_2S}}(P) =
\begin{cases}
X_{\mathrm{H_2S,deep}}, & P \ge P_{\mathrm{break}}, \\[6pt]
X_{\mathrm{H_2S,deep}}
\left(\dfrac{P}{P_{\mathrm{break}}}\right)^{\alpha_{\mathrm{H_2S}}}, & P < P_{\mathrm{break}},
\end{cases}
\]
where $X_{\mathrm{H_2S,deep}}$ denotes the deep, well-mixed mole fraction, $P_{\mathrm{break}}$ marks the pressure at which photochemical depletion begins, and $\alpha_{\mathrm{H_2S}}$ is a free power-law slope that governs the rate of depletion at lower pressures. We model the SO$_2$ VMR with a constant with pressure profile, given the non-detection of SO$_2$ in \citet{Fu2024}, we are not anticipating this to significantly impact our results. 

We therefore run five retrievals with the following assumptions:
\begin{enumerate}
    \item Chemical equilibrium following the same setup as seen in Section \ref{setup}.
    \item Same as above, plus the H$_2$S parameterisation.
    \item Elemental quenching following the same setup as seen in Section \ref{setup}.
    \item Same as above, plus the H$_2$S parameterisation.
    \item A free chemistry retrieval,
\end{enumerate}
and present the results of the disequilibrium and chemical equilibrium models in Table~\ref{tab:Retrieval_Results}, and the results of the free chemistry retrieval in Table \ref{tab:free_chemistry_guillot_results}. We plot the best-fitting spectra in Figure \ref{fig:Final_Plot} and the retrieved molecular volume mixing ratios in Figure \ref{fig:VMR_Plot}.

We find that the model setup that best describes the data (the largest Bayesian Evidence) is one which assumes disequilibrium chemistry with a ln(Z) = $571.8 \pm 0.036$. Comparing this to the chemical equilibrium retrieval, which has a ln(Z) = $564.5 \pm 0.108$, we find a log Bayes factor of $\Delta$ ln(Z) = 7.3 in favour of quenching of carbon-bearing molecules. We note that the quenching parameter for nitrogen-bearing molecules is unconstrained. The quenching is found to be deep in the atmosphere, with a $\log P_{\mathrm{q,C}}\,(\mathrm{bar})$ = $1.711^{+0.193}_{-0.281}$ compared to a chemical equilibrium model with the H$_2$S, this would suggest strong vertical mixing in the atmosphere.

If we now compare our models with and without the H$_2$S parameterisation, we find in the chemical equilibrium regime, there is a Bayes factor of 5 in favour of the model with the H$_2$S parameterisation. Though the same evidence is not seen between the disequilibrium models. Given that the disequilibrium model and the chemical equilibrium model are indistinguishable within the error of the observations (see Figure \ref{fig:Final_Plot}), and produce similar Bayesian evidence, we can say that these two assumptions produce two plausible descriptions of the atmosphere. The degeneracy can be broken by observations with NIRISS/SOSS. We show in Figure \ref{fig:Final_Plot} that these two models deviate by over 50 ppm in this region. The new bright mode of NIRISS/SOSS should be able to observe HD 189733b.

We demonstrate that we can begin to constrain the photochemically active regions of exoplanet atmospheres with JWST. We find that the photochemically active region of HD 189733b is log($P_{\mathrm{break}}$) = -3.25$^{+0.42}_{-0.51}$, consistent with predictions of photochemical models \citep{Tsai2021}. 

The retrieved photochemical depletion level of \ce{H2S} provides insights into the vertical structure of the atmosphere. First, it offers the first tentative constraint on the vertical gradient of an atmospheric gas in a hot Jupiter atmosphere \footnote{We note this has been done for ultra hot Jupiters, determining the pressure levels where the dissociation of H$_2$O occurs \citep{Evans-Soma2025}}. Second, $P_{\mathrm{break}}$ marks the pressure level at which the atmospheric composition begins to be heavily influenced by photochemistry. Based on model predictions, sulfur in this region primarily exists in atomic form \citep{Tsai2021, Tsai2023,Gruijter2025}. However, since the electronic transitions of atomic sulfur occur in the UV and visible, rather than the infrared, the location of $P_{\mathrm{break}}$ provides an indirect way of inferring atomic sulfur. Finally, our preliminary tests show that for atmospheres with spectral constraints on \ce{H2S} and \ce{SO2}, our parameterisation can potentially probe the transition layer between the vertically quenched region and the photochemical region, where sulfur is converted from \ce{H2S} into S, SH, \ce{SO2}, etc.

Our analysis finds the logVMR(H$_2$S) = $-3.21^{+0.18}_{-0.15}$ below the $P_{\mathrm{break}}$, whereas our free-chemistry retrieval finds the logVMR(H$_2$S) = $-3.71^{+0.31}_{-0.55}$, which is just outside of the 1-$\sigma$ upper limit of our deep abundance. We can conclude that the free-chemistry retrieval is biased by the vertical gradient of H$_2$S above $P_{\mathrm{break}}$, and the H$_2$S parameterisation provides a better constraint to the bulk sulfur inventory in the deep layers.

We compare the retrieved TP profiles and VMRs for three cases in Figure \ref{fig:VMR_Plot}: the disequilibrium retrieval,  chemical equilibrium retrieval that includes our H$_2$S parameterisation, and the free retrieval analysis. This helps to explain the difference in inferred metallicity between the chemical equilibrium assumption and our disequilibrium parameterisation. The retrieved VMRs from the free retrieval agree, to 1-$\sigma$, with the chemical equilibrium retrieval. Whereas the VMRs of CO, CO2, and H2O are larger in the disequilibrium retrieval, with a significant increase in CO that is incompatible with the free retrieval. The disequilibrium model needs to increase the temperature in the deep atmosphere to ensure the metallicity is high enough to match the H$_2$S abundance (which is only in chemical equilibrium), which therefore requires the quenching of CH$_4$ deep in the atmosphere to ensure it is depleted, as there is no evidence of CH$_4$ in the observation. This deep quenching results in the elevated VMRs of CO$_2$, CO, and H$_2$O.

\begin{figure*}
    \centering
    \includegraphics[width=\linewidth]{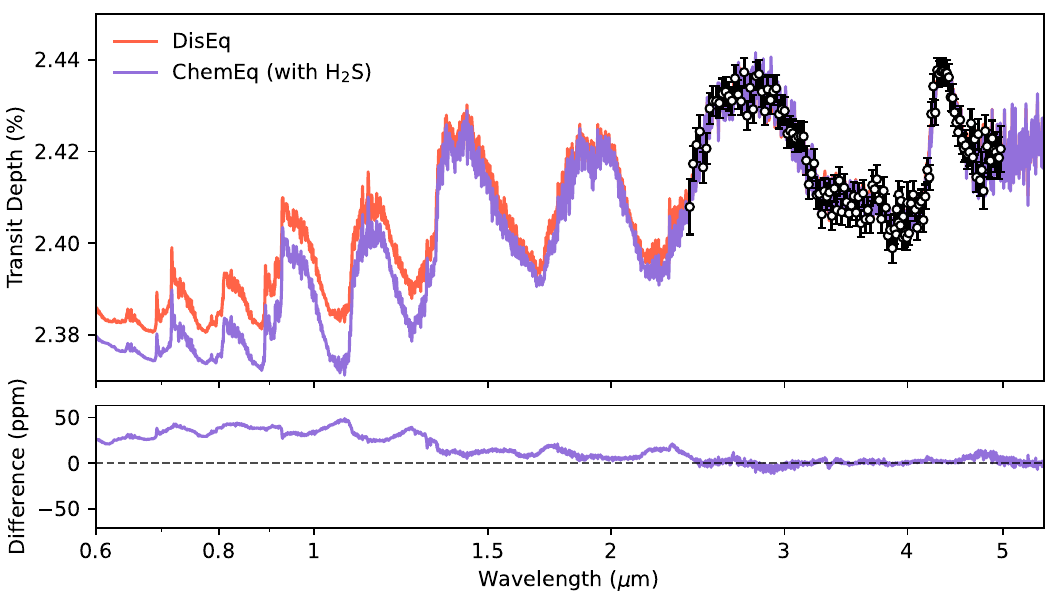}
    \caption{Comparison of the two models that best describe the observations. In red, we present the model where we have used our disequilibrium parameterisation described in Section \ref{sec:diseq_param}, and in purple, we show the model where we have assumed chemical equilibrium and used the H$_2$S parameterisation as described in Section \ref{sec:h2s_param}. The bottom panel shows the difference between the models. }
    \label{fig:Final_Plot}
\end{figure*}

\begin{figure*}
\centering
\includegraphics[width=\linewidth]{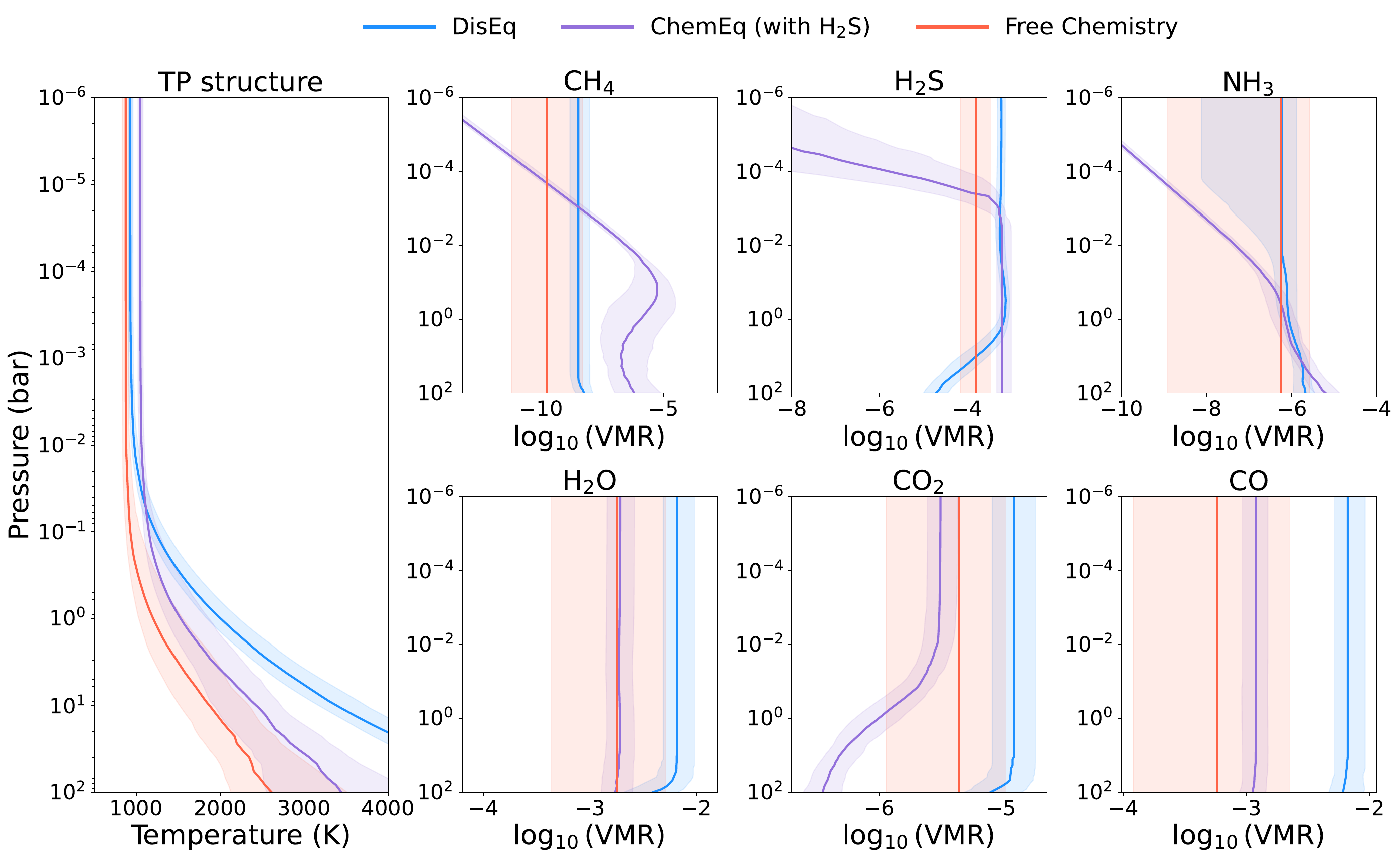}
\caption{Panels showing the TP profiles and volume mixing ratios of the molecules modelled in the disequilibrium (blue), chemical equilibrium with H$_2$S parameterisation, and free chemistry (red) analyses. Each profile has a 1-$\sigma$ envelope.}
\label{fig:VMR_Plot}
\end{figure*}

\begin{table*}
\centering
\begin{tabular}{lcccc}
\hline
Parameter & DisEq (with H$_2$S) & DisEq & ChemEq (with H$_2$S) & ChemEq \\
\hline
C/O & $0.407^{+0.114}_{-0.116}$ & $0.252^{+0.082}_{-0.061}$ & $0.345^{+0.107}_{-0.102}$ & $0.204^{+0.214}_{-0.185}$ \\
$$[M/H]$$ & $1.088^{+0.190}_{-0.486}$ & $1.631^{+0.086}_{-0.088}$ & $0.607^{+0.084}_{-0.063}$ & $0.726^{+0.717}_{-0.136}$ \\
$\log P_{\mathrm{q,C}}\,[\mathrm{bar}]$ & $-0.719^{+0.864}_{-3.665}$ & $1.711^{+0.193}_{-0.281}$ & - & - \\
$\log P_{\mathrm{q,N}}\,[\mathrm{bar}]$ & $-1.918^{+2.360}_{-2.485}$ & $-1.106^{+2.176}_{-3.028}$ & - & - \\
$\log X_{\mathrm{H_2S,deep}}$ & $-3.202^{+0.130}_{-0.136}$ & - & $-3.211^{+0.177}_{-0.145}$ & - \\
$\log P_{\mathrm{break,H_2S}}\,[\mathrm{bar}]$ & $-4.172^{+0.887}_{-1.033}$ & - & $-3.250^{+0.417}_{-0.507}$ & - \\
$\log\mathrm{(P_{\mathrm{top}})}\;(\mathrm{bar})$ & $0.033^{+1.191}_{-1.218}$ & $0.167^{+1.157}_{-1.335}$ & $-0.062^{+1.262}_{-1.327}$ & $-2.203^{+2.251}_{-0.130}$ \\
\hline
$\chi^2_\nu$ & $1.038$ & $1.023$ & $1.046$ & $1.181$ \\
$\ln Z$ & $569.4 \pm 0.061$ & $571.8 \pm 0.036$ & $569.5 \pm 0.036$ & $564.5 \pm 0.108$ \\
\hline
\end{tabular}%
\caption{Summary of retrieved results of the NIRCam observations of HD 189733b \citep{Fu2024}. We compare our disequilibrium parameterisation as described in Section \ref{sec:diseq_param} to the chemical equilibrium assumption. The simulations that have "with H$_2$S" include the H$_2$S parameterisation described in Section \ref{sec:h2s_param}}
\label{tab:Retrieval_Results}
\end{table*}

\section{Conclusions}
\label{Conclusions}
In this study, we have investigated how disequilibrium retrievals can be implemented and interpreted in the era of JWST. Using the photochemical model \texttt{VULCAN}, we modelled five atmospheric compositions for the hot Jupiter HD\,189733\,b: one assuming chemical equilibrium, and four assuming varying strengths of vertical mixing with $K_{\mathrm{zz}} = 10^6$, $10^8$, $10^{10}$, and $10^{12}$\,cm$^2$\,s$^{-1}$, respectively. These models were used to generate synthetic JWST transmission spectra for two commonly used instrument modes: NIRISS/SOSS and NIRSpec/G395H.

We developed a retrieval technique in which the quench pressures of carbon- and nitrogen-bearing species are treated as free parameters. This approach allows the chemical equilibrium abundances computed by \texttt{FastChem~3.0} to be truncated at the quench pressure, yielding vertically uniform abundances at lower pressures and thereby simulating the effects of transport-induced disequilibrium.

We applied our disequilibrium retrieval framework to synthetic JWST observations and found that, in the majority of cases, it outperforms the standard chemical equilibrium assumption when vertical mixing is present. With the inclusion of just two additional free parameters, the quench pressures for carbon- and nitrogen-bearing species, our approach enables accurate characterisation of the atmospheric composition of hot Jupiters that depart from thermochemical equilibrium. We extend this to individual molecules and demonstrate comparable retrieval performance compared to the elemental quenching.

We demonstrate that the quench pressures of carbon- and nitrogen-bearing species must be retrieved independently, as these species quench at different pressure levels. Our retrieval framework is capable of correctly recovering the input quenching pressure for each species, enabling more detailed probing of the underlying chemical processes occurring in exoplanetary atmospheres.

Finally, we applied our retrieval framework to recently published JWST observations of HD\,189733\,b and find that the atmosphere is best described by our disequilibrium framework, or a chemical equilibrium model, similar to those adopted by \citet{Fu2024} and \citet{Zhang2025}. Our best-fitting model is slightly different, however, as it incorporates a new parameterisation for the vertical profile of H$_2$S. Our retrieval yields a carbon-to-oxygen ratio of C/O = 0.345$^{+0.107}_{-0.102}$, which is bounded and does not tend to the lower limit. This is in contrast to the results of \citet{Fu2024}, which find from their grid retrieval a C/O = 0.12$^{+0.02}_{-0.01}$. We do obtain a similar C/O ratio when we do not parameterise the H$_2$S abundance. We find a [M/H] = 0.607$^{+0.084}_{-0.063}$, translating to 2-3.5$\times$ solar, which is consistent with \citet{Fu2024}. We demonstrate that by parameterising the VMR of H$_2$S with a pressure break and slope, it is possible to probe the photochemically active region of the atmosphere. We tentatively find evidence for this, leading to the first tentative evidence of the constraint on the photochemically active region in an exoplanet atmosphere.

We recommend that future retrieval studies of exoplanet atmospheres that assume chemical equilibrium consider including quench pressure parameters for key molecular species. This addition enables a more physically motivated characterisation of atmospheric composition and provides a means to account for the unknown strength of vertical mixing processes.

\section*{Acknowledgements}
The authors thank Dr Erik Meier Valdes for feedback that improved the manuscript. We thank the anonymous reviewer for their time and feedback; their comments enabled a deeper analysis and a clearer presentation of the results.
J.T. was supported by the Glasstone Benefaction, University of Oxford (Violette and Samuel Glasstone Research Fellowships in Science 2024). 

This work used the DiRAC Data Intensive service (DIaL2 / DIaL [*]) at the University of Leicester, managed by the University of Leicester Research Computing Service on behalf of the STFC DiRAC HPC Facility (\url{www.dirac.ac.uk}). The DiRAC service at Leicester was funded by BEIS, UKRI and STFC capital funding and STFC operations grants. DiRAC is part of the UKRI Digital Research Infrastructure.
\section*{Data Availability}
The data generated in this study are available on request.



\bibliographystyle{mnras}
\bibliography{bib} 



\appendix
\begin{table}
\centering
\begin{tabular}{lc}
\hline
Parameter & Value (median $\pm 1\sigma$) \\
\hline
$\log Z$ (evidence) & $567.58 \pm 0.04$ \\
Global reduced $\chi^2$ & $0.981$ \\
\hline

\multicolumn{2}{c}{Retrieved Molecular Abundances} \\
\hline
$\mathrm{H_2O}$ & $-2.66^{+0.40}_{-0.75}$ \\
$\mathrm{CO_2}$ & $-5.32^{+0.37}_{-0.66}$ \\
$\mathrm{CO}$ & $-3.16^{+0.51}_{-0.84}$ \\
$\mathrm{CH_4}$ & $-9.88^{+1.35}_{-1.31}$ \\
$\mathrm{HCN}$ & $-7.94^{+1.32}_{-2.46}$ \\
$\mathrm{NH_3}$ & $-6.35^{+0.80}_{-3.11}$ \\
$\mathrm{SO_2}$ & $-9.21^{+1.67}_{-1.70}$ \\
$\mathrm{H_2S}$ & $-3.71^{+0.31}_{-0.55}$ \\
\hline

\multicolumn{2}{c}{TP Profile Parameters} \\
\hline
$\log \kappa_{\mathrm{IR}}$ & $-3.36^{+0.45}_{-0.39}$ \\
$\log \gamma_1$ & $-2.09^{+1.18}_{-1.17}$ \\
$\log \gamma_2$ & $-2.03^{+1.20}_{-1.17}$ \\
$T_{\mathrm{irr}}$ (K) & $1021.31^{+62.29}_{-54.70}$ \\
$\alpha$ & $0.48^{+0.31}_{-0.30}$ \\
\hline

\multicolumn{2}{c}{Cloud and Scaling Parameters} \\
\hline
$\log P_{\mathrm{top}}$ & $0.27^{+1.01}_{-1.08}$ \\
Haze & $-0.69^{+2.98}_{-2.69}$ \\
Power & $5.50^{+2.75}_{-2.90}$ \\
$R_{\mathrm{scale}}$ & $0.9863^{+0.0020}_{-0.0027}$ \\
\hline
\end{tabular}
\caption{Retrieved parameters for the free-chemistry retrieval.}
\label{tab:free_chemistry_guillot_results}
\end{table}

\bsp	
\label{lastpage}
\end{document}